\documentstyle [12pt,epsfig]{article} 
\makeatletter
\@addtoreset{equation}{section}
\makeatother

\newcommand{\ba}{\begin{eqnarray}}
\newcommand{\ea}{\end{eqnarray}}

\begin{document}
\newcommand{\BS}{\bigskip}
\newcommand{\SECTION}[1]{\BS{\large\section{\bf #1}}}
\newcommand{\SUBSECTION}[1]{\BS{\large\subsection{\bf #1}}}
\newcommand{\SUBSUBSECTION}[1]{\BS{\large\subsubsection{\bf #1}}}

\begin{titlepage}
\begin{center}
\vspace*{2cm}
{\large \bf On the Relationship of Quantum Mechanics to Classical Electromagnetism
  and Classical Relativistic Mechanics }
\vspace*{1.5cm}
\end{center}
\begin{center}
{\bf J.H.Field }
\end{center}
\begin{center}
{ 
D\'{e}partement de Physique Nucl\'{e}aire et Corpusculaire
 Universit\'{e} de Gen\`{e}ve . 24, quai Ernest-Ansermet
 CH-1211 Gen\`{e}ve 4.}
\end{center}
\begin{center}
{e-mail; john.field@cern.ch}
\end{center}
\vspace*{2cm}
\begin{abstract}
 Some connections between quantum mechanics and classical
  physics are explored.  The 
Planck-Einstein and De Broglie relations, the wavefunction and its
probabilistic interpretation, the Canonical Commutation Relations and
the Maxwell--Lorentz Equation may be understood in a simple way by
  comparing classical electromagnetism and the photonic description of 
light provided by classical relativistic kinematics. The method used
may be described as `inverse correspondence' since quantum phenomena
 become apparent on considering the low photon number density limit of
classical electromagnetism. Generalisation to massive particles leads
 to the Klein--Gordon and Schr\"{o}dinger Equations.
 The difference between the quantum wavefunction of the photon and
 a classical electromagnetic
 wave is discussed in some detail.
\end{abstract}
\vspace*{1cm}
{\it Keywords}; Classical Electromagnetism, Relativistic Kinematics,
 Quantum Mechanics.
\newline
\vspace*{1cm}
 \par \underline{PACS 03.65.-w, 03.50.De, 03.30.+p}
 \par {\it To be published in European Journal of Physics}
\end{titlepage} 

\SECTION{\bf{Introduction}}
The traditional way to teach quantum mechanics (QM) is to introduce first the
non-relativistic theory. On the one hand there is the Schr\"{o}dinger 
Equation~\cite{Schrod}
 which resembles the other partial differential equations
 of mathematical physics, and like them has certain
 solutions, depending on the physical system under consideration
and boundary conditions. On the other hand, there is the 
physical interpretation of the wavefunction via Born's probabilistic
 postulate~\cite{Born1} and the
Heisenberg Uncertainty Relations~\cite{Heis}. It is in this
second part that the conceptual difficulty of the subject resides
and the borderline between physics and philosophy appears to be poorly
defined.
 The beginning student may have difficulty bridging the gap between the
 mathematical aspect (represented by
the Schr\"{o}dinger Equation, and its solutions, 
where many parallels may be found in classical physics)
and the physical interpretation
of QM.
\par It is suggested here that an alternative approach is possible, one that 
 follows closely (and even, in a sense to be explained later, {\it predates})
 the historical development of the subject. Planck was led to introduce the
 constant of action, $h$, by considering the properties of electromagnetic radiation in 
 thermal equilibrium~\cite{Planck1}, and the first example of `wave particle 
 duality' was Einstein's introduction of the light quantum
  concept~\cite{Ein1}. In both cases the
 system considered was ultra-relativistic
  \footnote{This terminology has two meanings in the present paper. Firstly, the photon velocity
   is evidently not negligible, as compared to the velocity of light, as is
  assumed for massive particles in non-relativistic quantum mechanics, and secondly, 
  that space and time are treated on an equal footing in all equations.}.
  The appearence of the constant
 $h$ in the non-relativistic Schr\"{o}dinger Equation, may then appear to be
 somewhat mysterious. The first idea for the 
 approach to QM adopted in this paper came from a reading of 
  Feynman's popular book `QED'~\cite{Feyn1}, where the fundamental
  concepts of QM are explained in terms of the interactions of 
  photons and electrons. Feynman's book presents many examples drawn 
  from physical optics, a subject treated in most textbooks
  in terms of the classical wave theory of light.  Contrary to 
 the traditional approach then, QM will be introduced here by way of the fully
 relativistic theory of free photons. The physical interpretation will then
 be shown to arise naturally by requiring consistency of this photonic
 (particle) description with the `wave' description supplied by 
 Classical Electromagnetism (CEM).
 \par A first step in this direction has been taken
 in an earlier paper~\cite{Field}
  by the present author in which the Lorentz Transformation (LT) was 
  derived on the basis of a few simple postulates, without reference to
  light or CEM. The kinematics of the LT was developed to show that 
  consistency with CEM requires Einstein's `light signals'
  to be interpreted as massless
  (or almost massless) particles, the `light quantum', discovered
   by Einstein, but in an, apparently, entirely different context.
   Thus, in the approach of Reference~\cite{Field}, the existence of photons
   is deduced from the correctness of Einstein's second postulate of
    Special Relativity: the constancy of the velocity of light, or alternatively,
     Einstein's second postulate, can be seen as a consequence of the light
    quantum concept.
   The comparison between the particle and wave aspects of light is 
   further developed in the present paper, and is shown to lead in a direct way to
   many of the fundamental concepts of QM. The  theory is 
    then generalised to particles
    of non-zero mass and the Klein Gordon and 
    Schr\"{o}dinger Equations are derived. The different physical interpretations
    of the wavefunction of QM and the electromagnetic waves of CEM are also
    discussed in connection with the photonic nature of light.
    \par An effort has been made throughout this paper to develop the subject
   in a logical manner, stating everywhere the assumptions that are made, 
   and building later results on ones established earlier. Still, classical 
   physics remains different from quantum physics and less fundamental. In the
   appropriate limit, the former may be derived from the latter, but the inverse
   procedure is not completely possible. Since the present paper works from
   known results of classical physics towards quantum physics there must then
   be, of necessity, a discontinuity in the argument. It will be  clearly pointed
   out just where this jump between the classical and quantum worlds is made, and
   where a new concept, not derivable from those used in the preceding arguments, must
   be introduced. Later, a logical bridge (without introducing any new concepts)
   will be built back from quantum mechanics to CEM.

    \par A related `historically correct' introduction to QM has been proposed
    in a recent paper~\cite{GM} that emphasised rather the relation of QM to
    thermodynamics and relativity, but still focused on the photonic 
    description of light. A brief comment on this approach is given in the 
    last Section of this paper.  
       
 \SECTION{\bf{Planck's Constant and the De Broglie Momentum-Wavelength
 Relation}}
  As first proposed by Minkowski~\cite{Mink}, to any physical object, O,
  of Newtonian mass, $m$, may be associated an energy-momentum 4-vector, $P$,
  defined as:
  \begin{equation}
 P \equiv m \frac{dX}{d \tau} \equiv (\frac{E}{c}; p_x,p_y,p_z)
\end{equation}
 where the space-time 4-vector $X$, is defined as: 
\begin{equation}
 X \equiv (ct;x,y,z) 
\end{equation}
and $\tau$ is the time  observed in the rest frame of O (its proper time).
 If the inertial frame S' is moving with uniform velocity $\beta c$ relative to
 the frame S along the common $x$, $x'$ axis, and $Oy$, $Oy'$ are also parallel,
 then the 4-vectors $P$, $P'$ as observed in S, S' are related by the LT
 equations:
\begin{eqnarray}
 p_x' & = & \gamma (p_x-\beta p_t)    \\
 p_y' & = & p_y \\
 p_z' & = & p_z \\ 
 p_t' & = & \gamma (p_t-\beta p_x )
\end{eqnarray}
where
\[ \gamma \equiv \frac{1}{\sqrt{1-\beta^2}},~~~~~~ p_t \equiv \frac{E}{c} \]
As discussed in Reference~\cite{Field}, $P$, is well-defined in the limit
$ m \rightarrow 0$, so that an energy-momentum 4-vector may also be
associated with the massless photon. In this case the 4-vector of a photon
of energy $E_{\gamma}$ moving in the $x$, $y$ plane, in a direction making an angle
$\phi$ with the x-axis in the frame S is:
\begin{equation}
P_{\gamma} = (\frac{E_{\gamma}}{c}; \frac{E_{\gamma}}{c} \cos \phi,
   \frac{E_{\gamma}}{c} \sin \phi, 0)
\end{equation}
A plane electomagnetic wave, which will here be associated with a large number of photons, each
having the same 4-vector, $P_{\gamma}$, was discussed by Einstein in his 
original paper on Special Relativity~\cite{Ein2}. 
  By considering the LT of
  space-time co-ordinates and electric and magnetic field components, the following
 equations were obtained for the transformation of the frequency, $\nu$, and the total energy,
 $E_T$, of a plane electromagnetic wave\footnote{More details of this derivation
  are given in Section 7 below}: 
\begin{eqnarray}
\nu' & = & \nu \gamma (1- \beta \cos \phi )  \\ 
E_T' & = & E_T \gamma (1- \beta \cos \phi )
\end{eqnarray}
$\nu$, $E_T$ refer to S and $ \nu'$, $E_T'$ to S'. Using
 Eqns.(2.6),(2.7) the energies of the photons 
that constitute the `electromagnetic wave' transform according to:
\begin{equation}
E'_{\gamma} = E_{\gamma} \gamma (1- \beta \cos \phi )
\end{equation}
If $n_{\gamma}$ is the total number of photons in the `light complex' considered
 by Einstein then $ E_T = n_{\gamma} E_{\gamma}$, and Eqn.(2.9)
  follows trivially from Eqn.(2.10).
 It also follows from Eqns.(2.8) and (2.10) that:
\begin{equation}
\frac{E_{\gamma}}{\nu} =  \frac{E'_{\gamma}}{\nu'} = constant
\end{equation}    
Calling the Lorentz invariant constant in Eqn.(2.11), $h$, then gives:
\begin{equation}
E_{\gamma} = h \nu
\end{equation}
 So the constant $h$ is identical to Planck's constant. The existence of the quantum
  of action, $h$, is then seen to arise from the consistency between the relativistic
  kinematics of photons, considered to be massless particles, described by Eqns(2.3)-(2.6)
 and the relativistic Doppler effect for classical electromagnetic waves, Eqn(2.8). 
  Thus, in a 
  certain sense, a fundamental equation of 
  QM results here from the fusion of two areas of classical physics.
  \par It should be noted that the photon concept is essential in the derivation of
   Eqn.(2.12) since Eqn(2.10) describes the kinematics of a massless particle.
   The `quantum of energy' that must be introduced in the derivation of Planck's
   radiation formula specifies only the manner in which black-body radiation
   is emitted or absorbed, not how it propagates in space-time. As the fierce debate
   around the Bohr-Kramers-Slater proposal~\cite{BKS} showed, it was, at first,
    not at all evident that the `quantum of energy' could be naively identified 
    with a particle~\cite{RAC}.
 \par The equation (2.12), together with the physical interpretation of $E_{\gamma}$ 
  as the energy
 of a `light quantum' had already been discovered earlier in 1905 by Einstein~\cite{Ein1}.
 Given the apparent simplicity of the above argument, one may ask why Einstein did not 
 notice how his Eqns.(2.8),(2.9) may be interpreted in terms of light quanta, given his
 knowledge of Eqn.(2.12). He limited himself to the remark: 
 \par {\sf ` It is remarkable that the energy and frequency of a light complex vary with
 the state of motion of the observer in accordance with the same law '}
 \par Actually, however, Einstein did not know, in 1905, the LT equations
  for the energy-momentum
 4-vector (Eqns.(2.3)-(2.6)) which were first written down in the following year by
 Planck~\cite{Planck2}.
 Denoting the modulus of the 3-momentum vector of the photon by $p_{\gamma}$, it
 follows from Eqn.(2.7) that $E_{\gamma}=p_{\gamma}c$. Substituting
 this relation in (2.12) and 
  using the definition of wavelength $\lambda = c/ \nu$, gives the
  de Broglie relation~\cite{Brog}:
 \begin{equation}
p_{\gamma} = \frac{h}{\lambda}
\end{equation}
The equivalent equation relating momentum and frequency: 
 \begin{equation}
p_{\gamma} = \frac{h \nu}{c}
\end{equation}
was actually discovered by Einstein in 1909~\cite{Ein3}, though not explicitly written
down as in Eqn(2.14)\footnote{See the excellent scientific biography of Einstein by A.Pais
~\cite{Pais}
 for a full account of the reticence
 towards the light 
quantum concept displayed by the physics community in the early years the last century.
This was, in large part, due to the enormous respect paid at that time to CEM.
  That the `wave' properties of this theory :
reflection, refraction, interference and diffraction are indeed {\it consequences}
 of the quantum
mechanics of photons was not recognised until much later. See, for example, Reference~\cite{Feyn1} .
Indeed it is clear that Einstein himself, although the originator of the
quantum theory of light, regarded it rather as a phenomenological, heuristic, model
rather than a complete and fundamental theory.}.
  This was first done by Stark~\cite{Stark}.
Further discussion of this point, in connection with Reference~\cite{GM}, may be found in
 Section 7 below. 
\SECTION{\bf{The Photon Wave Function and the Born Probabilistic Interpretation}}
\par From the expression for the energy density of a plane electromagnetic wave:
\begin{equation}
\rho_W = \frac {\vec{E}^2+\vec{B}^2}{8 \pi}
\end{equation}
the photon interpretation gives immediately Poynting's formula for the energy flow
$F$ per unit area per unit of time:
\begin{equation}
F = c \rho_W 
\end{equation}
as well as the formula for the radiation pressure $P_{rad}$ of a plane wave at normal incidence
on a perfect reflector:
 \begin{equation}
 P_{rad} = 2 \rho_W 
\end{equation}
Each photon annihilated by, and each photon created by the electrons in the reflector 
transfers a momentum, $p_{\gamma}$, perpendicular to its surface. The number of photons incident, per
unit area, per unit time, is $F/E_{\gamma}$. The total momentum transferred, 
 per unit area, per unit time,
is then $p_{\gamma}(F/E_{\gamma}) = F/c =\rho_W$. Because of the perfectly reflecting nature of the mirror, no 
net energy is absorbed, so an equal number of photons are re-emitted , also transferring
$\rho_W$ units of momentum, per unit area, per unit time, hence Eqn.(3.3).
  The classical result for radiation pressure, obtained in this way, requires the incident and reflected
  photons to have the same energy. For optical photons interacting with the electrons in the conduction
  band of a metallic reflector, this is expected to be a good approximation, the momenta of the
  photons being taken up by the atomic lattice as a whole, with negligible recoil energy.
  In general however, the energies of incident and reflected photons are different. This change in
  energy for the case of Compton scattering of X-rays was crucial for the general acceptance 
  of the validity of the photon concept~\cite{RAC,BG,CS}.  
\par A plane electromagnetic wave of wavelength $\lambda$, moving in free space
  parallel to the positive x direction
in the frame S may be written as:
\begin{eqnarray}
 E_y & = & E_0 \exp \Phi \\ 
 H_z & = & H_0 \exp \Phi
\end{eqnarray}
where:
\[ \Phi = 2 \pi i \frac{ (x-ct)}{ \lambda},~~~ E_0 = H_0 = A \]
The time-averaged energy density per unit volume, $ \bar{\rho}_W $, is: 
\begin{equation}
\bar{\rho}_W = \frac {E_0^2}{8 \pi}= \frac {H_0^2}{8 \pi}= \frac {A^2}{8 \pi}
\end{equation}
Assuming now that the wave consists of a beam of photons of energy $h \nu$ moving parallel
to the positive $x$ direction, the average number density of photons, $ \bar{\rho}_{\gamma}$, in the
wave is  $\bar{\rho}_W /h \nu$, so that, from Eqn.(3.6):
\begin{equation}
\bar{\rho}_{\gamma} = \frac {A^2}{8 \pi h \nu}
\end{equation}
 \par This is the point at which the leap across the chasm separating the
  classical and quantum worlds is made, and where as previously mentioned, a
  new physical concept must be introduced. The use of a complex exponential
  to represent a classical electromagnetic wave, as in Eqns(3.4) and (3.5) 
  is quite conventional. The wave may be associated with either the real
  or the imaginary part. On the other hand, for the description of light
  in QM, the use of the complete complex exponential is {\it mandatory}.
  Some {\it a posteriori} justification for this will be given below,
  but here the complex exponential is introduced as a necessary, but
  {\it a priori} unjustified, assumption. The second crucial point is
   to use the definition of wavelength together with Eqns(2.12) and (2.13)
   to replace, in the complex exponential the `wave' parameter $\lambda$
    by the `particle'
    parameters $E_\gamma$ and $p_{\gamma}$. The parameter $c$ that
  describes both the speed of propagation of electromagnetic waves
  and the velocity of photons is a part of both descriptions. Because the 
   equations of QM then contain only particle parameters, they
  describe, at the most fundmental level the behaviour of 
  particles, not of waves. Thus the `complex exponential describing
  photons' is obtained: 
\begin{equation}
u_p \equiv u_0 \exp \frac{2 \pi i}{h}(p_{\gamma}x-E_{\gamma}t) =  \exp \frac{2 \pi i}{h}(P \cdotp X)
\end{equation}
where
\begin{equation}
u_0 \equiv \frac{A}{\sqrt{8 \pi E_{\gamma}}}
\end{equation}
 The function $u_p$ will now be called the  {\it wavefunction} of the photon. As will be
 shown below, many key concepts and equations of QM may be derived from its mathematical
  properties.  
 Now Eqn.(3.7) may be written as:
\begin{equation}
\bar{\rho}_{\gamma} = | u_p |^2 = u_0^2
\end{equation}
For the case of very weak electromagnetic fields such that $ \bar{\rho}_{\gamma} \ll 1$,
 {\it $ | u_p |^2 dV$ is the probability that {\it a} photon is in the volume $dV$}. Thus 
  Eqn.(3.10) expresses  the Born probabilistic interpretation~\cite{Born1}
 the wavefunction $u_p$. It is derived by consistency with the energy density (3.1)
  of a classical electromagnetic wave in the low photon density limit. Introducing
 now differential operators ${\cal P}_x$, $\cal E$ according to the definitions:
 \begin{eqnarray}
{\cal P}_x & \equiv & -i\frac{h}{2 \pi}\frac{\partial~}{\partial x}   \\
{\cal E}  & \equiv & i\frac{h}{2 \pi}\frac{\partial~}{\partial t}
\end{eqnarray}   
 it can be seen that  $u_p$ is an eigenfunction of ${\cal P}_x$, $\cal E$
 with the eigenvalues $p_{\gamma}$, $E_{\gamma}$ respectively:    
\begin{eqnarray}
{\cal P}_x  u_p & = & p_{\gamma} u_p   \\
{\cal E} u_p & = & E_{\gamma} u_p
\end{eqnarray}
 This is an example of the general property of quantum wavefunctions 
  that they are eigenfunctions of operators corresponding to conserved
  (time independent) physical quantities of the system that they
   describe.
In CEM the exponential functions in Eqns.(3.4),(3.5) may be replaced, without loss of generality,
by real sinusoidal functions (as done, for example, in Reference~\cite{Ein2}). The wave function $u_p$ of QM
derived by `inverse correspondence' from Eqns.(3.4),(3.5) necessarily contains, however, an
exponential function of complex argument. Only in this way can the eigenvalue equations
 (3.13) and (3.14), corresponding to the fixed momentum
  and energy, respectively, of the free photon, be realised.
 It may also be remarked that the difference in sign in the definitions of the operators
 ${\cal P}_x$ and ${\cal E}$ in Eqns(3.11) and (3.12) is a consequence of the 
 Minkowski metric of the four-vector product between the space-time and energy-momentum vectors
 in the complex exponential factor of Eqn(3.8).
  \par In the above discussion, the photon wave function (3.8) has the form of a plane wave.
   Strictly speaking, such a wave function (for massless or massive particles) cannot be interpreted
   probabilistically according to Born's ansatz as a {\it spatial} wavefunction since a photon
  with such a wavefunction has the same probability to be found at any value of $x$. Indeed the
   wave function of Eqn(3.8) is not square integrable and so cannot yield a normalised probability
   density. However, Eqn.(3.10) does give correctly the probability to find {\it a} photon in unit volume
   in the infinitely long, but very diffuse, beam of photons that is the low density limit 
   of an electromagnetic plane wave. A further complication in the case of a photon it that, due to
    its massless nature, it is possible neither to assign to it a conserved probability current,
    nor to assign a quantum mechanical operator to its spatial position~\cite{NW,Amr}. This has lead to
    much discussion in the literature and text books of the `non-localisability' of photons
    ~\cite{MW}. Whether this is a real problem, going beyond the non-localisability of any
    particle (massive or massless) with a plane wavefunction is debateable, but beyond the scope
    of the present paper. From the practical, experimental, viewpoint it is evident that photons
    can be spatially localised as well (or as badly) as any other type of elementary particle.
     \par Because of the symmetry between $P$ and $X$ in Eqn.(2.8) it is possible,
     in a purely formal manner, to define operators ${\cal X} \equiv -i(h/2\pi)\partial/\partial p_x$
      and  ${\cal T} \equiv i(h/2\pi)\partial/\partial E$ analagous to  ${\cal P}_x$ and  ${\cal E}$
      respectively. The function $u_p$ of Eqn.(3.8)
      is an eigenstate of ${\cal X}$ with eigenvalue $x$ and of ${\cal T}$ with eigenvalue
      $t$. Such operators are not used in conventional QM, and, indeed, have no obvious utility
      or physical significance. The usefulness of the operators ${\cal P}_x$ and  ${\cal E}$
      is a consequence of Newtons's First Law, true in both classical mechanics and QM.
      The photons in a classical electromagnetic plane wave have fixed energy and 
    momentum in classical mechanics and are in eigenstates of energy and momentum in QM.
    A physical object is at a fixed position (or in an eigenstate of the spatial
    position operator) only in its rest frame. Such a frame does not exist for a massless 
    particle, which explains the absence of the operator ${\cal X}$ for such particles
    mentioned above. Given the intrinsic mutability of time an `eigenstate of time' seems to be
    both a logical and physical impossiblity.

\SECTION{\bf{The Canonical Commutation Relations, and Heisenberg Uncertainty
 Relation. The Maxwell--Lorentz Equation}}

  If $f$ is an arbitary function of space-time, it follows from Eqn(3.11) that
\begin{equation}
 {\cal P}_x x f =  -\frac{ih}{2 \pi}\frac{\partial(xf)}{\partial x}=
   -\frac{ih}{2 \pi}f+x {\cal P}_x f
\end{equation}
Defining the commutator $[A,B]$ according to:
\begin{equation}
[A,B] \equiv AB-BA,
\end{equation}
 Eqn(4.1) yields the Canonical Commutation Relation
 for the operator ${\cal P}_x$:

\begin{equation}
 [x,{\cal P}_x]   = \frac{ih}{2 \pi} 
 \end{equation}
 In the same way, it may be shown from Eqn(3.12) that:  
\begin{equation}  
 [ {\cal E}, t]    =  \frac{ih}{2 \pi}
\end{equation}
 As is well known~\cite{Robertson},
 the space-time Heisenberg Uncertainty Relation:
 \begin{equation}
  \Delta x \Delta p \ge \frac{h}{4 \pi}
 \end{equation}
 may be derived from equation (4.3) without additional
 hypotheses. Similarly Eqn.(4.4) may be used to derive the energy-time Uncertainty Relation:
 \begin{equation}
  \Delta E \Delta t \ge \frac{h}{4 \pi}
 \end{equation}
  Otherwise the physical meaning of the energy-time Canonical Commutation Relation
  (4.4) is somewhat obscure. Unlike Eqn.(4.3) that was crucial for Heisenberg's
   first calculation of atomic transition rates~\cite{Heis1}, Eqn(4.4) as no evident
  physical application.
\par Repeated use of Eqns.(3.11) to (3.14) and the relation $E_{\gamma} = p_{\gamma}c $ gives:
\begin{equation}
(c^2 {\cal P}_x^2- {\cal E}^2) u_p = (\frac{hc}{2\pi})^2 \Box u_p = 0
\end{equation}
Here $\Box$ is the d'Alembertian operator in one spatial dimension:
\begin{equation}
 \Box \equiv \frac{1}{c^2}\frac{\partial^2~}{\partial t^2}-
\frac{\partial^2~}{\partial x^2}
\end{equation}
Hence, from Eqn.(4.7),  $u_p$  satisfies the Maxwell--Lorentz Equation:
\begin{equation}
   \Box u_p = 0
\end{equation}
\par In the above discussion, only plane polarised waves with electric vector parallel
to the y-axis have been considered. Photon polarisation effects are easily 
taken into account by repeating the above derivation for other choices of polarisation
of the plane electromagnetic wave that gives, on replacing $\lambda$ by $E_{\gamma}$, $p_{\gamma}$,
 the wave function of the photon.

\SECTION{\bf{The Klein--Gordon and  Schr\"{o}dinger Equations }}  
\par Noting that the phase of the exponential in Eqn.(3.8) is a Lorentz invariant
quantity, being simply proportional to the 4-vector product of the space-time and 
momentum-energy vectors, that is equally valid for 
massless and massive particles, the above considerations are easily generalised to the case
of the latter.
Using instead of $E_{\gamma}=p_{\gamma} c$ the general energy-momentum relation for massive particles:
 \begin{equation}
 E^2 = p^2c^2+m^2c^4 
\end{equation}
together with Eqns.(3.11)-(3.14)
 the Maxwell--Lorentz Equation (4.9) generalises to the  
 Klein Gordon--Equation :
 \begin{equation}
 \left[(\frac{h}{2 \pi})^2 \Box + m^2 c^2 \right] u_p = 0
\end{equation}
In a similar way, the non-relativistic development of $E$ according to :
\begin{eqnarray}
 E=(p^2 c^2+ m^2 c^4)^\frac{1}{2} &  = &  m c^2 (1+\frac{p^2}{2 m^2 c^2} +... ) \\
 & \simeq & m c^2+\frac{p^2}{2m} = m c^2 + T^{(N)}  
\end{eqnarray} 
 together
with Eqns.(3.11)-(3.14) yields the  free particle Schr\"{o}dinger Equation:
 \begin{equation}
 \left [\frac{1}{2m}(\frac{h}{2\pi})^2 \frac{\partial^2~}{\partial x^2} +
  T^{(N)} \right] u_p^{(s)} = 0
\end{equation}
where 
\begin{equation}
 u_p^{(s)} \equiv u_p \exp[\frac{2 \pi i m c^2 t}{h}] = u_0 \exp [\frac{2 \pi i }{h}(px-T^{(N)}t)]
\end{equation}
and $T^{(N)}$ is the Newtonian kinetic energy $p^2/2m$. Here the differential operator
in Eqn.(3.12) corresponds to the Newtonian kinetic energy rather than the relativistic
energy $E$, and $u_p^{(s)}$ is an eigenstate of the $T^{(N)}$: 
 \begin{equation}
{\cal E} u_p^{(s)}  = i \frac{h}{2 \pi} \frac{\partial u_p^{(s)}}{\partial t} = 
  T^{(N)} u_p^{(s)}
\end{equation}
For the case of a particle interacting with a conservative, non-relativistic, potential, $V(x,t)$, Eqn.(5.5) 
still holds, after replacing $u_p^{(s)}$ with the general Schr\"{o}dinger  wave function $\psi^{(s)}$, and
using energy conservation to  write: 
\begin{equation}
  T^{(N)} = E^{(N)} - V(x,t)
\end{equation} 
where $E^{(N)}$ is the total Newtonian energy. Hence, using Eqn.(5.7), the one dimensional
 Schr\"{o}dinger Equation~\cite{Schrod} for an interacting particle is:
 \begin{equation}
 \left[ \frac{1}{2 m}(\frac{h}{2 \pi})^2 \frac{\partial^2~}{\partial x^2} +
 i\frac{h}{2 \pi} \frac{\partial~}{\partial t} -V(x,t)  \right] \psi^{(s)} = 0
\end{equation}
The differential operator in Eqn.(3.12) here corresponds to the total Newtonian energy
 $E^{(N)}$ defined in Eqn.(5.8).
 
\SECTION{\bf{Wavefunctions of Free and Confined Photons and Their Relation to
  Classical Electromagnetic Waves}} 
\par  The wavefunction describing a free photon, Eqn(3.8), can be used to construct both
  Stationary wavefunctions where a photon is confined to a limited spatial region, and 
  classical electromagnetic waves. In the latter case however, it will be seen that the physical
  interpretation becomes quite different to that of the quantum wavefunction.

\par Introducing the notation:
\begin{eqnarray}
 u_p^{(+)} & \equiv & u_0 \exp \frac{2 \pi i}{h}(p_{\gamma}x-E_{\gamma}t)  =  u_p \\
 u_p^{(-)} & \equiv & u_0 \exp \frac{2 \pi i}{h}(-p_{\gamma}x-E_{\gamma}t)
 \end{eqnarray}
 the Stationary wave function $\chi_E^{SW}$ is defined by the relation :
\begin{equation}
 \chi_E^{SW} \equiv \frac{ u_p^{(+)}+ u_p^{(-)}}{2} = u_0 \cos \frac{2 \pi p_{\gamma} x}{h} \exp \left[
 -\frac{2 \pi i E_{\gamma} t}{h} \right] 
\end{equation}       
As indicated by the subscript, $\chi_E^{SW}$ is an eigenfunction of the photon energy $E_{\gamma}$.
It is not, however, an eigenfunction of the photon momentum $p$. Eqn(6.3) shows that a photon
 in this state has equal amplitudes for photons with $p_x = \pm|p_{\gamma}|$. Such a wave function would
  describe, for example, a photon in a cavity
with perfect reflectors, perpendicular to the $x$-axis at
 $x = \pm(2n+1)h/2p$ where n is any integer. This Stationary wavefunction is the photonic
 analogue of an atomic bound state wavefunctions; for example, that of an electron bound 
 in a Hydrogen atom, that is also in an eigenstate of energy, but not of momentum.
 \par The second example is the wave function $\chi^{CPW}$, where the superscript CPW stands for 
 Classical Progressive Wave. This is defined as :
\begin{equation}
 \chi^{CPW} \equiv \frac{ u_p+ u_p^*}{2} = {\cal R}{\it e} (u_p) = 
  u_0 \cos \frac{2\pi}{h}(p_{\gamma}x-E_{\gamma}t)
  =   u_0 \cos \frac{2\pi(x- ct)}{\lambda}
\end{equation}
 The function $\chi^{CPW}$, with the replacements $p_{\gamma} \rightarrow h/\lambda$, 
 $E_{\gamma} \rightarrow h c /\lambda$, as in the last member
  of Eqn(6.4) yields a typical plane wave solution of Maxwell's Equations in CEM.
 This equation is then the bridge back across the chasm from the quantum to
  the classical world that was leapt, in the opposite sense,
  by the introduction of Eqn(3.8). Just as the quantum wavefunction is only a meaningful physical
  concept in the limit of very low photon density, then, as will now be 
  discussed, the function $\chi^{CPW}$ is meaningful only in the opposite limit of very
  high photon density.
 $\chi^{CPW}$ is an eigenfunction of neither $E_{\gamma}$ nor $p_{\gamma}$ and is a real,
  not a complex 
  function, as are the wavefunctions of a free or confined photon in Eqns(3.8) and (6.3)
  respectively. This illustrates the essential difference beween the classical electromagnetic
 ($\cal E \cal M$)  wave
 $\chi^{CPW}$  and the wave function $u_p$ of QM. It is common, in text books, to write classical
 $\cal E \cal M$ waves in just the exponential form (3.4),(3.5) with the understanding
 that either the real or $i$ times the imaginary part may be identified with a real 
 sinusoidal wave. The electric and magnetic field strengths in CEM are, of course, real
 quantities. Indeed, if a suitably orientated antenna is placed at any position in the
 wave $\chi^{CPW}$ then the time dependence of the induced voltage determines the phase
 of the electric vector at that position. Such a classical wave, produced for example
 by a dipole antenna, is always, then, represented by a real sinusoidal 
 function, although, for mathematical convenience only, the wave may
  be described by the real (or $i$ times the imaginary) part of a complex exponential. 
  The time average of $(\chi^{CPW})^2$, at any position, is 
  half the mean photon density, $\bar{\rho}_{\gamma}$
 which is related to the classical mean energy density $\bar{\rho}_W$ by
 $ \bar{\rho}_W = h \nu \bar{\rho}_{\gamma}$.  
 In a typical situation where the physics is well represented by CEM, the value of
 $\bar{\rho}_{\gamma} \Delta V$, where $\Delta V$ is the volume element
  of interest, is much larger than unity. For example, consider a radio transmitter of frequency
 1 MHz and a power of 1 kW with a short dipole antenna. The mean $\cal E \cal M$ field energy density at a distance of 
 10 km in a direction normal to the antenna is $\bar{\rho}_W = 3.9 \times 10^{-15} J/m^3$.
 The photon energy is $6.6 \times 10^{-28} J$, yielding a photon density of 
  $\bar{\rho}_{\gamma} = 5.9 \times 10^{12}/m^3$. If the wave is detected by an aerial of 
  length 1 $m$ and circular cross-section of 5 $mm$ diameter orientated parallel to the
   antenna, then in the minimum time of $ \simeq 1 \mu sec$ necessary to detect, classically, 
  the wave (i.e. to measure its amplitude, frequency and phase) the flux of photons 
  crossing the antenna is $\simeq 10^{13}$. Some small fraction of these photons will be 
  absorbed by conduction electrons in the aerial to provide the alternating current detected
  in the receiver. Note that the photons detected by the aerial are localised in a volume 
  with a typical spatial dimension much smaller than the wavelength (300 m) of the
  corresponding classical $\cal E \cal M$ wave. 
  It is clear that the probability that {\it a} photon will be found in
  the space-time volume appropriate to a classical measurement of the $\cal E \cal M$ wave
  is always unity.  The quantity $\bar{\rho}_W$ is the classical energy density 
  of the $\cal E \cal M$ field. However, $\bar{\rho}_{\gamma} \Delta V$ 
 has no probabilistic QM interpretation. 
  \par The relation of QM and CEM has been discussed in a similar way to above by 
   Sakurai~\cite{Sakurai}.
   A suggested condition for the validity of CEM was that the average energy density must be
  much larger than the scale of quantum fluctuations in the $\cal E \cal M$ field. This leads
  to the condition: 
  \begin{equation}
   \bar{\rho}_{\gamma} \gg \left(\frac{2 \pi}{\lambda}\right)^3
   \end{equation}
   Thus the number of photons in the volume $(\lambda/2 \pi)^3$ must be much larger than unity.
   In the example discussed above, where $\lambda/2 \pi  = 47.7$ m, the number of photons
   in the volume  $(\lambda/2 \pi)^3$ at the position of the aerial is $6.4 \times 10^{17}$, so
    that the condition (6.5) is, indeed,  well respected. In fact, the aerial would have to
    be at a distance $d$ from the transmitter such that:
  \[ \left(\frac{d}{10~km}\right) =  6.4 \times 10^{17} \]
  or  $d = 8 \times 10^9$ km (about 53 times the distance from the Earth to the Sun)
   before the condition (6.5) for the validity of CEM breaks down.
   \par The classical limit can also be discussed by use of second quantised field theory.
  As shown by Sakurai~\cite{Sakurai} the commutation relations between photon creation
   and annihilation operators can be used to derive a `number phase' Uncertainty
   Relation:
 \begin{equation}
    \Delta N \Delta \phi \ge 1
 \end{equation}  
 showing that precise determination of the classical phase, $ \phi$, requires a large uncertainty in the
  photon occupation number, $N$. It is clear from (6.6) that for a single photon ($N = 1$,
   $\Delta N = 0$) the classical phase is completely uncertain. Finally, Sakurai~\cite{Sakurai}
   also discusses how Eqn.(6.6) can be interpreted in such a way as to yield
   a momentum-space Uncertainty Relation similar to Eqn.(4.5) for a localised $\cal E \cal M$ 
    wave-train.

\SECTION{\bf{Summary and Conclusions}}        
In this paper a method has been suggested for introducing the basic concepts of 
QM, not, as is done in the majority of text books on the subject, via the
Schr\"{o}dinger Equation and the non-relativistic theory, but rather by re-interpreting
the equations of CEM (a subject well-known by the majority of 
students confronting QM for the first time) to take into account the
existence of photons, massless particles described by classical relativistic kinematics.
 Historically, Planck's constant~\cite{Planck1} and the light quantum~\cite{Ein1} were discovered
  by considerations based on the properties of cavity radiation and the photoelectric effect,
  respectively.
 An alternative way of deriving the 
Planck-Einstein relation, Eqn.(2.12), is via the photonic description of light
in Special Relativity and CEM.
Einstein's second postulate in his Special Relativity paper~\cite{Ein2}
concerning the constancy of the velocity of light is a direct consequence~\cite{JMLL,Field} of the
LT, on the assumption that light consists of massless (or almost massless)
particles: photons. Comparing the LT of the frequency of a plane $\cal E \cal M$ wave
(Eqn.(2.8)) with that of the energy-momentum 4-vector of a photon, Eqn.(2.10),
then leads directly to the Planck-Einstein relation: $E_{\gamma} = h \nu$. 
\par  The wavefunction of the photon and its Born probabilistic interpretation may be introduced
  by considering the low photon density limit of a classical $\cal E \cal M$ wave. 
  The wavefunction is shown to be an eigenfunction of the quantum mechanical
  operators corresponding to the momentum and energy of the photon.
The properties of the wavefunction of a free photon are then shown to yield the Canonical Commutation
Relations of QM, and hence the Heisenberg Uncertainty Relation, as well as the Maxwell--Lorentz equation.
 Generalisation of the Maxwell--Lorentz equation
to the case of massive particles enables the derivation of the Klein--Gordon and  Schr\"{o}dinger 
Equations. The distinction between the quantum wavefunctions of free photons, Stationary
 photon wavefunctions and classical $\cal E \cal M$ waves is discussed. The classical wave appears in
 the high particle density limit of quantum mechanics. Classical $\cal E \cal M$ waves can
  always be represented
 by real functions, whereas quantum wavefunctions are, necessarily, complex in 
 order to represent eigenfunctions of the operators of conserved quantities.

  \par It should be emphasised that, just as a large part of the motivation
  of the present paper resulted from ideas suggested by a reading of Reference~\cite{Feyn1},
  Feynman's approach to relativistic quantum mechanics has been followed throughout this
  paper. In the more conventional quantum field theory of Dirac~\cite{Dirac} photons
  are described in terms of `second-quantised' electromagnetic field 
  operators. However, all physical results of this
  formalism are equivalent to those of the Feynman diagram approach~\cite{Feyn2} where
  only first-quantised (`c-number', in Dirac's terminology) wave functions,
  as discussed in the present paper, appear in the invariant amplitude
  for any physical process\footnote{ This approach is applied systematically 
  in Reference~\cite{HM}. Note especially the discussion of
   Gauge Theories on P313.}. It should also be stressed that, nowhere in the present
  paper, is the dynamics of photon production and destruction, described in second-quantised field 
  theory by creation and annihilation operators, discussed. The quantum mechanical principles follow
  solely from consideration of the kinematical properties of free photons. Indeed
  dynamics occurs in the paper only, under the guise of energy conservation, in the
  discussion of the Schr\"{o}dinger Equation in Section 5. 

 \par There is an appreciable overlap between the ideas presented in the present paper
  and those contained in a recent paper of Margaritondo~\cite{GM}. For example,
  the same simple derivation of the law of radiation pressure, using the photon
 concept, given above, was presented. The main emphasis of Reference~\cite{GM} was,
  however, on Einstein's original discovery of the light quantum concept
  by comparing the entropy of a dilute gas with that of cavity radiation
  with frequencies within the range of validity of Wien's Law. Einstein's
  remark concerning the similarity of the transformation laws of the 
  frequency of electromagnetic waves and of the energy of the electromagnetic
  field, quoted in Section 2 above, was also mentioned. Margaritondo suggests
  that Einstein may have (though tacitly) also deduced the existence of photons
  from this similarity, and this may have given him confidence to publish
  the `revolutionary' paper in which the existence of light quanta was actually
  proposed~\cite{Ein1}. To support this argument, Margaritondo surmises that
  Einstein interpreted his formula for the energy of an electromagnetic wave according
  to Eqn(2.10) above which describes the transformation of the energy of a photon.
  A careful reading of Einstein's Special Relativity paper~\cite{Ein2} shows, however,
   that although the transformation law of the `light complex' considered is indeed
    Eqn(2.9), which seems, to modern eyes, trivially related to Eqn(2.10), Einstein
    actually derived this relation by a completely different route that has no obvious
   connection with the photon concept. The total energy inside a spherical surface, drawn
   inside a plane electromagnetic wave, as seen by two different observers, one at rest
 and the other in motion, was considered. In Eqn(2.9), $E_T$ and $E_T'$ actually represent 
  the energy content of the electromagnetic field, within this spherical surface, as seen
   by the two observers. To arrive at Eqn(2.9), both the Loretntz transformation of
  the electric and magnetic fields as well as the length contraction of the sphere, as
  viewed by one of the observers, was taken into account. Thus the increase of the 
  energy due to the transformation of the electromagnetic fields was partially
  compensated by the length contraction effect. It seems unlikely that Einstein would
  have presented such an intricate derivation if he was aware that Eqn(2.9) may be almost
  trivially derived by use of the photon concept. In fact, as already pointed out in 
   Section 2 above, Einstein was not aware, in 1905, of the LT laws for the energy-momentum
   four-vectors, Eqns(2.3)-(2.6) above. Indeed, if Einstein had had a full
   understanding of both relativistic kinematics and the photon concept at this time, it
   would have been unnecessary to pose the constancy of the velocity of light as a
  separate axiom in Special Relativity, since it is a necessary consequence of the
   identification of light as massless particles~\cite{JMLL,Field}. In fact, as discussed in
  considerable detail in Reference~\cite{Pais}, Einstein's attitude towards the photon 
  concept (i.e. considering light to fundamentally {\it be} a particle) remained
  ambiguous until the end of his life. Indeed, he rather considered fields (physical
  entities existing continously in space-time) to be the most fundamental concepts in physics.
  In fact,
  Eqn(2.9) was derived from consideration of electromagnetic fields, not light quanta.
   It is not for nothing, then, that the `light quantum' paper describes, according to
   Einstein, not a `theory' but rather a: `... Heuristic Point of View Concerning the
  Production and Transformation of Light'.
  
\par Finally, the remark may be made, as previously pointed out by Feynman~\cite{Feyn1}
 and other
authors adopting a similar approach~\cite{JMLLFB},
 that the so called `classical wave theory of
light' developed in the early part of the 19th Century by Young, Fresnel and others
{\it is} QM as it applies to photons interacting with matter. Similarly Maxwell's theory of
CEM is most economically regarded as simply the limit of QM when the number of photons
involved in a physical measurement becomes very large. Indeed, in this paper, QM has been understood
 by `inverse correspondence', considering the opposite (small photon number) limit of CEM.
 Thus experiments performed with light by physicists during the last century, and even earlier,
 {\it were} QM experiments, now interpreted via the wavefunctions of QM, but then in terms
  of `light waves'.
 The latter may be derived from the former by the replacements: $p \rightarrow h/\lambda$
  and $E \rightarrow h/\nu$. The `particle' parameters $p,~E$ of the wave function are replaced by the
  `wave' parameters $\lambda,~\nu$, and $h$ disappears from the equations, 
  yielding the classical wave theory of light. The predictions
  for, say, interference and diffraction, remain however, fundamentally, those of QM.
  It is in this sense, as mentioned in the introduction, that the quantum theory of light {\it predates}
  the discoveries of Planck and Einstein. They actually discovered that light consists of
  particles, that Special Relativity theory shows to have kinematical properties
  reassuringly close to those of any other physical objects of classical physics.
  The essential and mysterious aspects of QM, as embodied in the wavefunction
  (superposition, interference) were already well known, in full mathematical detail,   
  almost a hundred years earlier!
 \par \underline{Acknowledgements}
 \newline I should like specially to thank V.Telegdi for his reading of an early
 version of this paper, and for his critical comments. The remarks on
 the manuscript by P.Achard and B.Echenard, that helped to improve
 the presentation, are also gratefully acknowledged.  
  Finally I thank an anonymous referee for a comment that led me to clarify
  the distinction between the concepts of relativistic quantum mechanics in
  space-time, as formulated by Feynman, and those of traditional quantum field 
  theory.

\pagebreak
 
\end{document}